\documentclass[aps, prd, twocolumn, lengthcheck, superscriptaddress, showpacs, letterpaper, nofootinbib]{revtex4-1}

\usepackage{amsmath,amssymb}
\usepackage{graphicx,bm}
\usepackage{slashed}
\usepackage{dcolumn}
\usepackage{epstopdf}
\usepackage{ulem} 
\usepackage[usenames]{color}
\usepackage{mathrsfs}

\allowdisplaybreaks

\def\la{\langle}\def\ra{\rangle}
\def\be{\begin{eqnarray}}\def\ee{\end{eqnarray}}
\def\lsim{\mathrel{\rlap{\lower3pt\hbox{\hskip1pt$\sim$}}
     \raise1pt\hbox{$<$}}} 
\def\gsim{\mathrel{\rlap{\lower3pt\hbox{\hskip1pt$\sim$}}
     \raise1pt\hbox{$>$}}}
\def\bi{\bibitem}

\begin{document}

\title{Topology change and nuclear symmetry energy in compact-star matter}

\author{Xiang-Hai Liu}

\affiliation{College of Physics, Jilin University, Changchun,
130012, China}

\author{Yong-Liang Ma}
\email{yongliangma@jlu.edu.cn}
\affiliation{Center for Theoretical Physics and College of Physics, Jilin University, Changchun,
130012, China}

\author{Mannque Rho}
\email{mannque.rho@cea.fr}
\affiliation{Institut de Physique Th\'eorique,
CEA Saclay, 91191 Gif-sur-Yvette c\'edex, France }

\date{\today}
\begin{abstract}
We show that the cusp-like structure in the nuclear symmetry energy, a consequence of topology change,  visible in the skyrmion lattice description of dense matter is extremely robust. It is present in the Skyrme model -- with the pion field only -- and is left unscathed by massive degrees of freedom such as the vector mesons $\rho$ and $\omega$ and also the scalar meson introduced as a dilaton. It leads to the emergence of parity-doubling at the skyrmion-half-skyrmion  transition and impacts on the nuclear symmetry energy ``wilderness" landscape, weeding out roughly half of the wilderness. It is shown that  a pseudo-conformal sound velocity arises at a precocious density at which the cusp is formed. It is suggested that the topology change, being robust, could impact on the tidal deformability observed in gravity waves

\end{abstract}


\maketitle


\section{introduction}
\label{sec:intro}
The nuclear symmetry energy $E_{sym} (n,\alpha)$ at baryonic density $n$ that appears quadratically in  $\alpha=(N-P)/(N+P)$ in the energy per nucleon $E (n,\alpha)$  where $P$ ($N$) is the number of protons (neutrons)  in many nucleon systems given by
\be
E(n, \alpha) & = & E(n, \alpha = 0) + E_{\rm sym}(n)\alpha^2 + O(\alpha^4) + \cdots ,
\label{eq:defEsym}
\ee
plays the most important role in the equation of state (EoS)) for compact-stars~\cite{Li:2008gp,Steiner:2004fi}. In standard nuclear physics approaches (SNPAs) anchored on the effective density functional (EDF) such as the Skyrme potential, relativistic mean field (RMF) and varieties thereof (\cite{EDF} for a recent review) and on standard chiral perturbation (S$\chi$PT) expansion  up to manageable chiral order (e.g., \cite{holt-rho-weise}), equipped with a certain number of parameters fit to available empirical data, the $E(n,\alpha)$ can be more or less reliably determined in the vicinity of the nuclear matter equilibrium density $n_0\sim 0.16$ fm$^{-3}$. It has also been extended, with a rather broad range of uncertainty, up to slightly above $n_0$ from heavy-ion collision experiments.  Thus the nuclear symmetry energy (NSE)  is fairly well determined up to  $n_0$ in SNPAs  although its slope in density remains still somewhat uncertain.

\begin{figure}[h]
\begin{center}
\includegraphics[width=7cm]{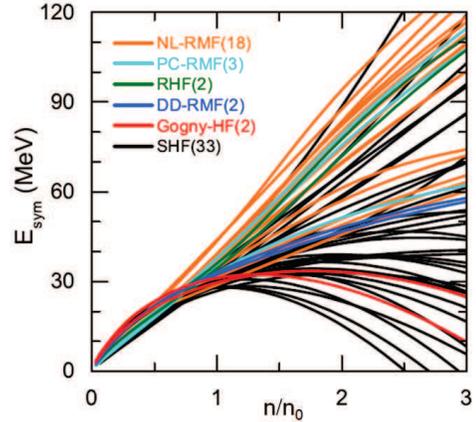}
 \end{center}
 \vskip -0.7cm
\caption{A sample of predicted $S=E_{sym}$ by models -- ``symmetry energy landscape" -- that are fit to the matter properties up to $n_0$ and extrapolated to higher densities, showing the wilderness at high densities. Copied from Ref.~\cite{chen-wilderness}.}
\label{mess}
\end{figure}

Going beyond $n_0$, particularly above $2n_0$ which is crucial for massive compact stars, however, is totally unknown as one can see  in Fig. \ref{mess} (from \cite{chen-wilderness})  which summarizes the total wilderness in the NSE landscape beyond $n_0$ predicted by hundreds of models in SNPA.
More recent analyses~\cite{zhang-li} that take into account up-to-date information including the LIGO-Virgo gravitational waves from coalescing  neutron stars~\cite{GW170817} do clear up the situation somewhat, but the wilderness largely remains.  What this implies is that the structure of the EoS established at $n_0$ cannot be extrapolated reliably  beyond $\sim n_0$.

There can be two possible causes for the breakdown of  straightforward extrapolation.
(A) One plausible cause for this state of matter is that beyond $\sim n_0$ other degrees of freedom than the hadrons that figure in the SNPAs could be playing an influential  role. (B) Another possibility  is that even if hadrons are {\it the} relevant degrees of freedom in the range of density involved, the approximation used may be breaking down at some high density. For instance, the standard chiral perturbative approach  with the nucleons and pions as the relevant degrees of freedom can be reliable only if the energy/momentum scale involved is in a sense ``soft," which in baryonic matter involves a ``soft" Fermi momentum $k_F$ that  cannot be much bigger than chiral symmetry scale  $f_\pi$. One example is the currently popular three-body force for nuclear matter and nuclear structure in S$\chi$PT approaches, of which the short-range component involves the mass scale of an $\omega$ meson which is of higher scale than the typical cutoff figuring in standard effective chiral theories. Therefore the chiral expansion cannot be trusted at densities higher than, say, 2$n_0$. On the other hand, the Fermi-liquid structure of nuclear matter at $n_0$ which may be taken as a justification for RMF approaches~\cite{matsui} is an expansion in $1/\bar{N}$ around the Fermi-liquid fixed point~\cite{shankar} with
\be
\bar{N}=k_F/(\Lambda -k_F) \label{Nbar}
\ee
where $\Lambda$ is the cutoff on top of the Fermi surface.  The fixed point involved here is the fixed point in the Fermi sphere appropriate for the equilibrium state of nuclear matter at $n_0$. Now going to higher densities beyond $n_0$, the Fermi sphere could be modified by a change of the relevant degrees of freedom or a change in the structure of the relevant degrees of freedom.
The consequence is that either the Fermi-liquid fixed point present at $n_0$ could then become unstable or the system could develop a different Fermi-liquid structure with a different fixed point at high density. Either of  the two cases could invalidate the standard density expansion of the SNPAs. Unfortunately the only known trustful  QCD approaches currently available, i.e., perturbative QCD and lattice simulations, cannot access the density regime involved in compact stars and cannot guide the density expansion, thus accounting for the wilderness Fig.~\ref{mess}.

One way out of this conundrum was proposed in \cite{LPR} exploiting a topology change involving skyrmion structure of dense  baryonic matter, dual to (or mimicking) the baryon-quark continuity expected in QCD. It is the purpose of this paper to show that the notion of topology change at high density  is robust and leads to an important  change in the structure of high density matter.  It indeed gives in particular a unique parameter-free nuclear symmetry energy for $n>n_{1/2}$. If proven correct, this observation  would neatly clear up the wilderness in the NSE landscape.
\section{Cusp-like structure of nuclear  symmetry energy}
The principal idea we wish to develop is based on the notion that the skyrmion description of baryonic matter could give a hint in what way soliton description  can capture QCD in the large $N_c$ limit at high density~\cite{multifacet}. At that limit, the baryonic matter can be taken as a skyrmion crystal matter. Now it is found that at a high density, denoted as $n_{1/2}$ hereon,  a skyrmion with the topological number $B=1$ put on FCC crystal fractionalizes into a pair of half-skyrmions on CC with $B=1/2$. This follows from symmetry considerations~\cite{goldhaber-manton,kugler}. In the absence of fully quantized skyrmion matter, which remains currently an unsolved problem, one cannot say at what density this fractionalization takes place\footnote{A recent analysis~\cite{MR-PCM} gives the range $2\lsim n_{1/2}/n_0\lsim 4$. For the present discussion, we need not specify it and assume that this ``transition" takes place at $n_{1/2} > n_0$.}. It appears that this half-skyrmion structure is already present in the alpha particle $^4$He~\cite{battyeetal}. But the eight half-skyrmion configuration is less favored energetically than four-skyrmion configuration in $^4$He unlike the  matter at high density. The half-skyrmions  are not propagating objects because they are confined by  monopole~\cite{cho,Nitta}. What makes the difference between skyrmion matter and half-skyrmion matter on the crystal lattice, which is the most crucial ingredient in what follows, is that the bilinear quark condensate $\Sigma\equiv \la\bar{q}q\ra$ where $q$ is the light-quark field, when averaged over space, is non-zero in the former ($\bar{\Sigma}\neq 0$) with $f_\pi\neq 0$, but while locally nonzero, goes to zero when averaged in the latter ($\bar{\Sigma}=0$), with however non-vanishing $f_\pi$. This means that $\Sigma$ is not a bona-fide order parameter for chiral symmetry on the lattice.

What was shown in \cite{LPR} is the consequence of the topology change at $n_{1/2}$ on the nuclear symmetry energy in the Skyrme model~\cite{Skyrme} --  with the pion field only -- with the quadratic current algebra term plus the Skyrme quartic term\footnote{A scalar field was supplemented by means of the  conformal compensator field for dilaton. However the scalar plays a minor role in the symmetry energy, so will be ignored in this (qualitative) work.}.  The symmetry energy in the skyrmion description is given by the neutron matter collectively quantized. It is given in the large $N_c$ limit by the isospin moment of inertia ${\lambda_I}$ as
\be E_{sym}\propto 1/{\lambda_I} +O(1/N^2_c).
\ee
 A surprising result found in \cite{LPR} with the skyrmions put on crystal lattice was that $E_{sym}$ has a ``cusp" structure\footnote{We are unable to determine whether the ``cusp" has a true discontinuity in slope or an artifact of the crystal structure. In reality, this is of no consequence physics-wise because it is smoothed in nature by higher correlations as is observed in renormalization-group calculations going beyond the semi-classical approximation~\cite{PKLMR} and also by the chiral symmetry explicit breaking. Although we continue to refer to it as a cusp we will consider it as a cusp-like structure.} with the symmetry energy decreasing as density approaches $n_{1/2}$ from below and then increases as density increases away from $n_{1/2}$.  Now the major difference, as pointed out above,  between the skyrmion matter and the half-skyrmion matter is the global structure of the condensate $\Sigma$. Thus this cusp structure must reflect the change from $\bar{\Sigma}\neq 0$ to $\bar{\Sigma}=0$ at $n_{1/2}$.
\section{The cusp and  EFT Lagrangian }
We should stress here that this cusp structure is not an artifact of the crystal approximation. It turns out to play a crucial role in giving an effective field theory (EFT) Lagrangian that provides a realistic equation of state (EoS) for compact-star matter

We first see how the cusp structure dictates the change of parameters in the effective EFT Lagrangian at a density corresponding to  $n_{1/2}$.  To develop the argument,  we start by assuming that the result of skyrmion lattice at density near $n_{1/2}$ is robust. We will then show in what follows that this assumption can be validated.

Consider an effective (continuum) Lagrangian in which light-quark baryons and pions are coupled in hidden local symmetric way to the vector-meson fields $\rho$ and $\omega$. It has the usual chiral symmetry, with the HLS Lagrangian being gauge-equivalent to the non-linear sigma model.  It is known that the nuclear tensor force plays a dominant role in the nuclear symmetry energy, so we look at the tensor forces given by the exchange of a $\pi$ and a $\rho$. Phrased in terms of nuclear forces, which is a well-established procedure to do effective field theory in nuclear matter, a good approximation to the symmetry energy can be obtained by the closure-sum approximation~\cite{closure-sum},
$E_{sym}\approx C \la |V_T^2|\ra/\bar{E}$ where $C$ is a known constant, $V_T=V_T^\pi +V_T^\rho$ is the radial part of the sum of the pionic tensor force and $\rho$ tensor force and $\bar{E}\approx 200$ MeV is the average energy of the state to which the tensor force predominantly connects from the ground state. There are also contributions from other components of nuclear forces but we ignore them at the level we are considering, which is effectively semiclassical.  A characteristic feature of the $\pi$ and $\rho$ tensor forces is that they enter with an opposite sign. So the pion tensor,  dominant in the range relevant in the nuclear matrix element, say, $r\gsim 0.7$ fm,   with the  contributions for $r<0.7$ fm screened by short-range correlations,   gets canceled progressively by the $\rho$ tensor at increasing density. In medium, the effective $\rho$ mass drops  with density while the pion mass remains more or less unaffected. Therefore as the density increases and the quark  (or dilaton) condensate decreases, the $\rho$ tensor force strength increases,   thereby diminishing the net tensor force, making it eventually go to zero at $n\sim 3n_0$ -- for certain set parameters determined at $n_0$ --  if naively extrapolated to higher densities.  This would make the symmetry energy effectively go to zero at $\sim 3n_0$.  This  would then give an $E_{sym}$ belonging to the class of  the curves in Fig.~\ref{mess} that turn over and drop to zero near $3n_0$. This feature is clearly at variance with the cusp-like structure given by the topology change.

What happens is that at a density commensurate with the topology change density, $n_{1/2}$,  the $\rho$ tensor undergoes a drastic change.  This is because at $\sim n_{1/2}$ the hidden gauge coupling $g$, constant up to $n_{1/2}$,  starts dropping rapidly at increasing density so that it goes,  driven by renormalization group flow, to zero arriving at what is called ``vector manifestation (VM)" fixed point which is thought to be coincident with chiral restoration~\cite{HY:PR}.  Since the $\rho$ mass is given by the KSRF formula to all orders of loop corrections~\cite{HY:PR}, the $\rho$ mass,  going as $m_\rho^2\sim f_\pi^2 g^2$, tends to zero independently of what $f_\pi$ is at the VM fixed point. This rapid drop in the gauge coupling makes the $\rho$ tensor strongly suppressed at $n_{1/2}$ making the $\pi$ tensor take over completely.  This exactly reproduces the cusp predicted by the skyrmion crystal model.

The closure-sum result with the tensor force only -- and no kinetic energy term -- is of course an approximation. The EFT Lagrangian implemented with the parameter changes at $n_{1/2}$ mentioned above and  other channels than the tensor force  taken into account, treated in RG-flow approach~\cite{PKLMR}, smoothens the cusp, but it retains the behavior for density above $n_{1/2}$ given by the crystal calculation.
\section{Vector mesons as hidden local symmetric fields}
\subsection{HLS Lagrangian}
So far we have treated the role of the pion field  in the crystal structure. The generic structure of the cusp is dictated by the pion field that carries the topology but there is a strong indication that massive vector fields  play a very important role in the structure of baryonic matter. This aspect is amply illustrated in the hadron/nuclear section in the recent volume of collected papers on skyrmions in condensed matter, hadron/nuclear physics and string theory~\cite{multifacet}. In the field of hadron/nuclear physics, the subject ranges from finite nuclei to infinite nuclear matter, including compact-star matter. Surprisingly remarkable is the role the vector mesons, in particular the $\rho$ meson, play in the skyrmion description, not only for binding energies  but also for cluster structure of finite nuclei~\cite{sutcliffe-cluster}. It is thus an extremely interesting and relevant question to ask what the vector mesons do to the cusp-like structure of the symmetry energy. Here we consider how the presence of the $\rho$ and $\omega$ mesons influences the symmetry energy when treated on crystal lattice.

How the vector mesons are incorporated in chiral symmetric Lagrangian has been extensively discussed in the literature. In the context of this work, we adhere to the approach adopted most recently in \cite{MR-PCM}, namely in implementing hidden local symmetry (HLS)~\cite{HY:PR}\footnote{It is perhaps worth reiterating the point which is not well understood in the field, particularly in connection with the property of vector mesons near the putative chiral restoration temperature in relativistic heavy ion collision. The key idea in exploiting hidden gauge symmetry for the cusp-like structure in $E_{sym}$  is the possible coincidence of vector manifestation with chiral symmetry restoration. As explained in \cite{MR-PCM}, local gauge symmetry is ``hidden" in the description of the $\rho$ meson only when the mass is identically zero in the framework of nonlinear sigma model in the chiral limit~\cite{hidden}.  Thus probing the ``vector manifestation" in dilepton processes (as wrongly identified as probing Brown-Rho scaling) requires being infinitesimally close to the vector manifestation fixed point. Within the experimental accuracy in the background of strong nuclear interactions, this point has been accessed in none of the experiments performed up to now.}  which is particularly powerful in dealing with dense matter.  As mentioned, the scalar dilaton field that figures importantly in \cite{PKLMR,MR-PCM} does not play significant role for the symmetry energy, so we leave it out.

We work with the hidden local symmetric Lagrangian for $N_f=2$ that we write down to $O(p^4)$ in the power counting that is generalized to include vector mesons.  There are 14 $O(p^4)$ terms. In general the 14 parameters are unknown but in the limit that the gravity/string duality holds, they can be fixed in terms of the two parameters,  $f_\pi$, the pion decay constant and $m_\rho$, the $\rho$ meson mass that figure in the holographic dual Lagrangian.   It will also turn out that the fourteen terms can be reduced to a good approximation to only four terms.

The HLS Lagrangian to $O(p^4)$ is written in three terms
\begin{eqnarray}
\mathcal{L}_{\rm HLS} & = & \mathcal{L}_{\rm (2)} +
\mathcal{L}_{\rm (4)} + \mathcal{L}_{\rm anom}
\label{eq:Lag_HLS}
\end{eqnarray}
where the subscript $(m)$ represents the chiral order $O(p^m)$, $m=2,4$ and ${\cal L}_{anom}$ is the anomaly term, which in the case of 2 flavors, is the homogeneous Wess-Zumino (hWZ) term.   In writing down the lengthy formulas, it makes it economical to use two Maurer-Cartan 1-forms
\begin{eqnarray}
\hat{\alpha}_{\parallel\mu} & = & \frac{1}{2i}(D_\mu \xi_R \cdot
\xi_R^\dag +
D_\mu \xi_L \cdot \xi_L^\dag), \nonumber\\
\hat{\alpha}_{\perp\mu} & = & \frac{1}{2i}(D_\mu \xi_R \cdot
\xi_R^\dag - D_\mu \xi_L \cdot \xi_L^\dag)
\ ,. \label{eq:1form}
\end{eqnarray}
Here $\xi_{L,R}$ are the chiral fields that figure in the pion field  $U(x)=\exp(i\pi^a\tau^a/f_\pi)=\xi_L^\dag(x)\xi_R(x) \in SU(2)_L\times SU(2)_R/SU(2)_{L+R}$, and
 the covariant derivative is defined as $D_\mu \xi_{R,L} = (\partial_\mu - i V_\mu)\xi_{R,L}$ with the hidden local field $V_\mu=\frac {g}{2} (\tau^a V^a_\mu + \omega_\mu)$.  We will take the vector mesons with $U(2)$ symmetry, although at high density the symmetry is found to be broken~\cite{PKLMR}. The symmetry breaking does not affect the discussion that follows.

 In terms of the 1-forms, the leading order $O(p^2)$ Lagrangian is
\begin{eqnarray}
\mathcal{L}_{\rm (2)} & = &
f_\pi^2 \,\mbox{Tr}\, \left( \hat{a}_{\perp\mu}
\hat{a}_{\perp}^{\mu} \right)
+ a f_\pi^2 \,\mbox{Tr}\, \left(\hat{a}_{\parallel\mu}
\hat{a}_{\parallel}^{\mu} \right) \nonumber\\
& &{}
- \frac{1}{2g^2} \mbox{Tr}\, \left( V_{\mu\nu}V^{\mu\nu} \right),
\label{eq:HLSp2}
\end{eqnarray}
where $f_\pi$ is the pion decay constant, $a$ is the parameter of the HLS,
$g$ is the hidden gauge  coupling constant, and $V_{\mu\nu} = \partial_\mu V_\nu-\partial_\nu V_\mu-i[V_\mu,V_\nu]$ is the field-strength tensor of field $V_\mu$. The chiral symmetry breaking quark mass term appears also at the same order that we have not written down.

Since we need the explicit form of the $O(p^4)$ terms for the analyses given below, we write $\mathcal{L}_{(4)}$ in two terms
\begin{eqnarray}
\mathcal{L}_{(4)y} & = &
y_1^{} \mbox{Tr} \Bigl[ \hat{\alpha}_{\perp\mu}^{} \hat{\alpha}_\perp^\mu
\hat{\alpha}_{\perp\nu}^{} \hat{\alpha}_\perp^\nu \Bigr]
+ y_2^{} \mbox{Tr} \Bigl[ \hat{\alpha}_{\perp\mu}^{} \hat{\alpha}_{\perp\nu}^{}
\hat{\alpha}^\mu_\perp \hat{\alpha}^\nu_\perp \Bigr] \nonumber\\
& &{} + y_3^{} \mbox{Tr}
\left[ \hat{\alpha}_{\parallel\mu}^{} \hat{\alpha}_\parallel^\mu
\hat{\alpha}_{\parallel\nu}^{} \hat{\alpha}_\parallel^\nu \right]
+ y_4^{} \mbox{Tr}
\left[ \hat{\alpha}_{\parallel\mu}^{} \hat{\alpha}_{\parallel\nu}^{}
\hat{\alpha}^\mu_\parallel \hat{\alpha}^\nu_\parallel \right]
\nonumber \\ && \mbox{}
+ y_5^{} \mbox{Tr}
\left[ \hat{\alpha}_{\perp\mu}^{} \hat{\alpha}_\perp^\mu
\hat{\alpha}_{\parallel\nu}^{} \hat{\alpha}_\parallel^\nu \right]
+ y_6^{} \mbox{Tr}
\left[ \hat{\alpha}_{\perp\mu}^{} \hat{\alpha}_{\perp\nu}^{}
\hat{\alpha}^\mu_\parallel \hat{\alpha}^\nu_\parallel \right] \nonumber\\
& &{} + y_7^{} \mbox{Tr}
\left[ \hat{\alpha}_{\perp\mu}^{} \hat{\alpha}_{\perp\nu}^{}
\hat{\alpha}^\nu_\parallel \hat{\alpha}^\mu_\parallel \right]
\nonumber \\ && \mbox{}
+ y_8^{} \left\{
\mbox{Tr} \left[ \hat{\alpha}_{\perp\mu}^{} \hat{\alpha}_\parallel^\mu
\hat{\alpha}_{\perp\nu}^{} \hat{\alpha}_\parallel^\nu \right]
+ \mbox{Tr} \left[ \hat{\alpha}_{\perp\mu}^{} \hat{\alpha}_{\parallel\nu}^{}
\hat{\alpha}_\perp^\nu \hat{\alpha}_\parallel^\mu \right] \right\} \nonumber\\
& &{} + y_9^{} \mbox{Tr}
\left[ \hat{\alpha}_{\perp\mu}^{} \hat{\alpha}_{\parallel\nu}^{}
\hat{\alpha}^\mu_\perp \hat{\alpha}^\nu_\parallel \right],
\\
\mathcal{L}_{(4)z} & = &
i z_4^{} \mbox{Tr}
\Bigl[ V_{\mu\nu} \hat{\alpha}_\perp^\mu \hat{\alpha}_\perp^\nu \Bigr]
+ i z_5^{} \mbox{Tr}
\left[ V_{\mu\nu} \hat{\alpha}_\parallel^\mu \hat{\alpha}_\parallel^\nu \right].
\label{eq:HLSp4}
\end{eqnarray}

Finally the anomalous term, which is also of $O(p^4)$, consists of three terms,
\begin{eqnarray}
\mathcal{L}_{\rm anom} & = & \frac{N_c}{16\pi^2}
\sum_{i=1}^3 c_i \mathcal{L}_i ,
\label{eq:HLShWZ}
\end{eqnarray}
where
\begin{subequations}
\begin{eqnarray}
\mathcal{L}_1 & = & i \, \mbox{Tr}
\bigl[ \hat{\alpha}_{\rm L}^3 \hat{\alpha}_{\rm R}^{}
 - \hat{\alpha}_{\rm R}^3 \hat{\alpha}_{\rm L}^{} \bigr], \\
\mathcal{L}_2 & = & i \, \mbox{Tr}
\bigl[ \hat{\alpha}_{\rm L}^{} \hat{\alpha}_{\rm R}^{}
\hat{\alpha}_{\rm L}^{} \hat{\alpha}_{\rm R}^{} \bigr]  ,  \\
\mathcal{L}_3 & = & \mbox{Tr}
\bigl[ F_{\rm V} \left( \hat{\alpha}_{\rm L}^{} \hat{\alpha}_{\rm R}^{}
 - \hat{\alpha}_{\rm R}^{} \hat{\alpha}_{\rm L}^{} \right) \bigr] ,
\end{eqnarray}
\end{subequations}
in the 1-form notation with
\begin{eqnarray}
\hat{\alpha}_{L}^{} &=& \hat{\alpha}_\parallel^{} - \hat{\alpha}_\perp^{},
\nonumber \\
\hat{\alpha}_{R}^{} &=& \hat{\alpha}_\parallel^{} + \hat{\alpha}_\perp^{},
\nonumber \\
F_V &=& dV - i V^2.
\end{eqnarray}
\subsection{Approximating HLS}
It is at present not feasible to do a fully systematic calculation to $O (p^4)$ in the way chiral effective field theories are performed in nuclear dynamics. There are too many parameters to fix  by theory and phenomenology. There is however one way to do the full $O(p^4)$ calculation if the parameters are extracted from a holographic QCD model, in particular, the Sakai-Sugimoto model~\cite{Sakai:2004cn}. The Sakai-Sugimoto model given in 5D has only two parameters related to the known quantities, the pion decay constant  and the $\rho$ mass. Reduced \`a la Klein-Kaluza from 5D to 4D in infinite towers of vector mesons, integrating out all higher members of the towers than the lowest vector mesons $\rho$ and $\omega$ can fix all the parameters of the Lagrangian (\ref{eq:Lag_HLS})~\cite{Harada:2010cn}. They are listed in Table \ref{table:LECs}~\cite{Ma:2012zm}.  Of course, although chirally symmetric, the Sakai-Sugimoto model is at best a rough approximation to QCD in light-quark systems\footnote{For instance, the Sakai-Sugimoto model may not have the correct ultraviolet completion of QCD.}, so it should be taken with a grain of salt.
\begin{table*}[t]
\caption{\label{table:LECs} Low energy constants of the HLS Lagrangian
at $O(p^4)$ with $a=2$ determined from Sakai-Sugimoto model given in Ref.~\cite{Ma:2012zm}.}
\begin{ruledtabular}
\centering
\begin{tabular}{clllllllll}
Parameter
& \, $y_1=-y_2$ &\, $y_3=-y_4$ & \, $y_5=2y_8=-y_9$ & \, $y_6=-(y_5+y_7)$ & \qquad $z_4$ &
\qquad $z_5$ & \qquad $c_1$ & \qquad $c_2$ & \qquad $c_3$ \\
\hline
Value & $- 0.001096$ & $ - 0.002830$ &\;\;\; $-0.015917$ &\;\;\; $+0.013712$ &
$0.010795$ & $-0.007325$ & $+0.381653$ & $-0.129602$ & $0.767374$ \\
\end{tabular}
\end{ruledtabular}
\end{table*}
This Lagrangian with all the parameters so determined will be referred to as ``HLS $(\pi,\rho,\omega)$".

We consider four possible approximations to the full HLS Lagrangian including one that seems to mysteriously work well.
\begin{enumerate}
\item The first is the Skyrme Lagrangian which corresponds to ${\cal L}_{(2)}$ with the vector mesons integrated out. We refer to this as ``HLS ($\pi$)."
\item The second is ${\cal L}_{(2)}$ with all $O(p^4)$ terms ignored. In the absence of the hWZ term, the $\omega$ decouples from $\pi$ and $\rho$. This is referred to as ``HLS ($\pi$, $\rho$)."
\item The third is ${\cal L}_{\rm min}\equiv {\cal L}_{(2)} + \frac 12 g\omega_\mu$ that we call HLS$(\pi,\rho,\omega)_{\rm min}$. The second term $\frac 12 g\omega_\mu$ is obtained from the hWZ Lagrangian when, reduced to two terms,  the equation of motion for the $\rho$ meson with the kinetic energy term dropped -- which amounts to taking $m_\rho\to \infty$ -- is substituted into it\footnote{As stressed later and illustrated below, this is a bad approximation given that  the VM fixed point is forsaken.}.
\item The fourth is a drastic simplification to  the full HLS, HLS$(\pi,\rho,\omega)$,   where all the $O(p^4)$ terms except for the hWZ term and $i z_4^{} \mbox{Tr}
\Bigl[ V_{\mu\nu} \hat{\alpha}_\perp^\mu \hat{\alpha}_\perp^\nu \Bigr]$ from $\mathcal{L}_{(4)z} $ are dropped. We shall call this ``HLS $(\pi,\rho,\omega)_{ \rm simp}$."  In the absence of the hWZ term, the $\omega$ field decouples. The resdulting  $O(p^2)$ Lagrangian turns out to work surprisingly well for iso-vector processes involving the pion and the $\rho$ meson, both in and out of nuclear matter. How this comes about remains a mystery~\cite{komargodski}.
\end{enumerate}

\section{Symmetry energy in skyrmion crystal}
\label{sec:SE}

We now turn to simulating the symmetry energy in the skyrmion crystal model.

As first proposed in \cite{LPR}, we obtain the symmetry energy by collective-quantizing the  pure neutron matter in skyrmion crystal~\cite{Klebanov}. Here we quantize the crystal as a whole object through a collective rotation in iso-space with the rotation angle $C(t)$ acting on the relevant chiral fields as
\begin{eqnarray}
\xi_c(\mathbf{x}) & \to & \xi(\mathbf{x}, x) = C(t) \xi_c(\mathbf{x}) C^\dagger(t),
\nonumber \\
V_{\mu,c}(\mathbf{x}) & \to & V_{\mu}(\mathbf{x},t) =  C(t) V_{\mu,c}(\mathbf{x}) C^\dagger(t) ,
\label{eq:rotation}
\end{eqnarray}
where the subindex ``$c$" means the static configuration with the lowest energy for a given crystal size $L$ and $C(t)$ is a time-dependent unitary $SU(2)$ matrix in isospace. We  define the angular velocity through ${\bm{\Omega}}$
 \begin{eqnarray}
\frac{i}{2}\bm{\tau}\cdot\bm{\Omega} & = & C^\dagger(t)\partial_0 C(t) .
\label{eq:defvelocity}
\end{eqnarray}
The energy of the $n$-nucleon system can be written as
\begin{eqnarray}
M_{\rm tot} & = & M_{\rm static} + \frac{1}{2}\lambda_{I}^{\rm tot} \bm{\Omega}^2.
\end{eqnarray}
By regarding the angular momentum in isospace, $\mathbf{J} = \delta M_{\rm tot}/\delta \bm{\Omega}$, as the isospin operator, one can write the total energy of the system  as
\begin{eqnarray}
M_{\rm tot} & = & n M_{\rm sol} + \frac{1}{2n \lambda_{I}}I^{\rm tot}(I^{\rm tot} + 1),
\end{eqnarray}
where $M_{\rm sol}$, $\lambda_{I}$ and $I^{\rm tot}$ are, respectively, the mass and moment of inertia of the single skyrmion in the system, and the total isospin of the $n$-nucleon.
Given that the $n$-nucleon system is taken a nearly pure neutron system,  $I^{\rm tot} \leq n/2$, to the leading order of $n$ for $n \to \infty$, the energy per baryon takes the form
\begin{eqnarray}
E & = & M_{\rm sol} + \frac{1}{8 \lambda_{I}}\alpha^2.
\end{eqnarray}
Thus the symmetry energy is
\begin{eqnarray}
E_{\rm sym} & = & \frac{1}{8 \lambda_{I}}.
\label{eq:LEsymLambda}
\end{eqnarray}

Putting HLS on crystal lattice is discussed extensively in \cite{PRV,MHLOPR} for static properties of the skyrmion matter. We simply take the formulas from these references.
The collective rotation~\eqref{eq:rotation} generates certain components of the vector-meson fields which are absent in the static case: The time component of $\rho$ meson field and the spatial component of $\omega$ meson field are excited~\cite{Meissner:1986js}. For example, in the quantization of the single skyrmion,
\begin{eqnarray}
\bm{\tau}\cdot \bm{\rho}^0(\bm{r},t) & = & \frac{2}{g}C(t)\bm{\tau}\cdot\left[\bm{\Omega}\xi_1(r) + \hat{\bm{r}}\xi_2(r)\right]C^\dagger(t), \nonumber\\
\bm{\omega}(\bm{r},t) & = & \frac{\phi(r)}{r}\bm{\Omega}\times \hat{\bm{r}}.
\end{eqnarray}
These components make the calculation more involved than what was done in ~\cite{LPR} in which only the pion and dilaton fields figured.

To parameterize the time component of $\rho$ meson field, we introduce the following Fourier-serious expanded quantity
\begin{eqnarray}
\bar{ \phi} _0^{\rho_ 0}  =  \sum_{abc} {\beta _{abc}^{\rho_0}} \cos \left( {\frac{{a\pi x}}{L}} \right)\cos \left( {\frac{{b\pi y}}{L}} \right)\cos \left( {\frac{{c\pi z}}{L}} \right)\\
\end{eqnarray}
with which the time component of $\rho$ meson field is given by
\begin{eqnarray}
V_0 & = & \frac{g}{2}\rho_0 \nonumber\\
& = & \frac{1}{1+ \bar{ \phi} _0^{\rho_ 0}}C(t)\left[\left(\bm{\bar{\phi}}^{\rho_0}\cdot\bm{\bar{\phi}}^{\rho_0}\right)\left(\bm{\tau}\cdot \bm{\Omega}\right) \right.\nonumber\\
& &\left. \qquad\qquad\qquad{} - \left(\bm{\bar{\phi}}^{\rho_0}\cdot\bm{\Omega}\right)\left(\bm{\tau}\cdot \bm{\bar{\phi}}^{\rho_0}\right) \right]C^\dagger(t).
\end{eqnarray}

The parameterization of the spatial component of omega meson field $\omega_i$ is more involved. Naively, one may write it as
\begin{eqnarray}
\bm{\omega} & = & \bm{\Omega}\times \bm{\hat{\omega}}
\end{eqnarray}
with $\bm{\hat{\omega}}$ being an iso-scalar three-dimensional vector and then Fourier-expand $\bm{\hat{\omega}}$ in the same form as $\bar{\phi}_i^\pi$ give in ~\cite{PRV}. If, for simplicity, one assumes rotational symmetry for the three-vector and takes $\bm{\Omega} = (1,0,0)$, then we can parameterize $\omega_i$ as
\begin{eqnarray}
\omega_1 & = & 0 , \\
\omega_2 & = & \sum_{lmn} \gamma _{lmn}^{\omega_2} \cos \left( {\frac{{l\pi x}}{L}} \right)\cos \left( {\frac{{m\pi y}}{L}} \right)\sin \left( {\frac{{n\pi z}}{L}} \right) , \\
\omega_3 & = & \sum_{lmu} \gamma _{lmn}^{\omega 3} \cos \left( {\frac{{l\pi x}}{L}} \right)\sin \left( {\frac{{m\pi y}}{L}} \right)\cos \left( {\frac{{n\pi z}}{L}} \right)  .
\end{eqnarray}
However if we substitute the above expansion into the Lagrangian to calculate the moment of inertia, we find the trivial solution $\gamma _{lmn}^{\omega_2} = \gamma _{lmn}^{\omega_3} = 0$. In order to have a nontrivial contribution to the moment of inertia as in the static case, we adopt to use, as the equation of motion of $\omega_i$,
\begin{equation}
{} - \partial_\beta \partial^\beta \omega^\alpha + \partial_\beta \partial^\alpha \omega^\beta = 2{g^2}f_\pi ^2\omega^\alpha + W^\alpha - \partial _\beta Z^{\beta\alpha } ,
\end{equation}
where $W^\alpha$ and $Z^{\beta\alpha }$ are functions of the pion and $\rho$ meson fields which can be derived from~\eqref{eq:Lag_HLS}. It follows that the Fourier coefficients $\gamma _{lmn}^{\omega_i}$ can be expressed in terms of those of $W_i$ and $Z_{ij}$, the pion and $\rho$ meson fields. This gives a nontrivial contribution from $\omega_i$.

To arrive at the moment of inertia, we first determine the Fourier coefficients of the static pion, $\rho$ and $\omega$ fields by minimizing the static energy of the skyrmion crystal at a certain size and then we substitute such static $\pi$, $\rho$ and $\omega$ field configurations and obtain the moment of inertia  by adjusting the Fourier coefficients of the excited fields such that the minimal value of the moment of inertia at that crystal size is obtained.  The numerical simulation exploited to obtain the results given below is straightforward, though involved with many terms. A brief description of the procedure is relegated to  Appendix.

\section{Results}
\label{sec:Num}

\begin{figure}[h]
\begin{center}
\includegraphics[width=8cm]{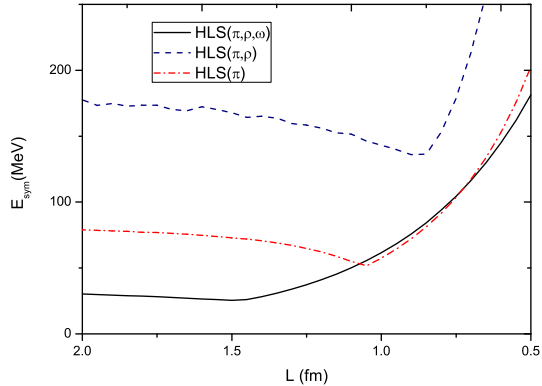}
\caption{The symmetry energy as a function of crystal size $L$. Here and in all the figures that
follow, $L$ decreases to the right, indicating increase in density. The numerical results for HLS$(\pi,\rho,\omega)$,  HLS$(\pi,\rho)$ and  HLS$(\pi)$ are presented by solid, dashed, and dash-dotted
lines, respectively. }.
\label{fig:EsymHLS}
\end{center}
\end{figure}

We first compare the results from HLS$(\pi)$, HLS$(\pi, \rho)$ and HLS$(\pi, \rho, \omega)$ to see the effect of the $\rho$ and $\omega$ mesons on the density dependence of the symmetry energy. We recall that HLS$(\pi)$ is actually the Skyrme model treated in \cite{LPR} with the dilaton incorporated.
We see from From Fig.~\ref{fig:EsymHLS} that all three Lagrangians produce a cusp. The position of the cusp is determined by the skyrmion-half-skyrmion changeover density $n_{1/2}$. This has to do with the global quark condensate $\bar{\Sigma}$ vanishing at that density with the emergence of parity doubling.

Note that while the location of the cusp density depends on the degrees of freedom, the existence of the cusp is generic,   dependent only on the pion field. It is the topology that is in action.

To see that the position of the cusp is locked to the topology change density $n_{1/2}$, we look at the expression of the moment of inertia from the model HLS$(\pi)$. That is
\begin{eqnarray}
\lambda_{I} & = & \frac{f_\pi^2}{6}\left\langle \left(4 - 2 \phi_0^2\right)\right\rangle + \cdots ,
\label{eq:MOISkyr}
\end{eqnarray}
where $\cdots$ stands for the contribution from the Skyrme term and $\langle \cdots \rangle$ indicates the space average of the quantity inside. Since with the decreasing of the crystal size, or increasing of the density, $\langle\phi_0^2\rangle={\bar{\Sigma}}^2$ decreases to zero and $1/\lambda_{I}$ decreases  going toward $n_{1/2}$.  What happens when density exceeds $n_{1/2}$ is  highly involved with the intervention of massive excitations such as the quartic term in the Skyrme model and the vector mesons in the HLS models, which account for the increase in the symmetry energy for $n>n_{1/2}$.

A remarkable observation to make here is the key role played by the vector mesons $\rho$ and $\omega$ at densities higher than their respective $n_{1/2}$. In terms of the structure of the tensor force in the closure approximation~\cite{LPR} discussed above, how the vector manifestation figures for $n> n_{1/2}$ in controlling the $\rho$ tensor force depends on an intricate interplay between the $\rho$-nucleon coupling and $\omega$-nucleon coupling (as discussed in \cite{PKLMR}).   The behavior of the symmetry energy above the topology change density clearly reflects this feature.

It is interesting  also that Fig.~\ref{fig:EsymHLS} shows the magnitude of the symmetry energy in HLS$(\pi,\rho)$ greater than that in HLS$(\pi)$. This is because, as in the matter-free space~\cite{Ma:2012zm}, for a given density, the moment of inertia in HLS$(\pi,\rho)$ is smaller than that in HLS$(\pi)$.

\begin{figure}[h]
\begin{center}
\includegraphics[width=8cm]{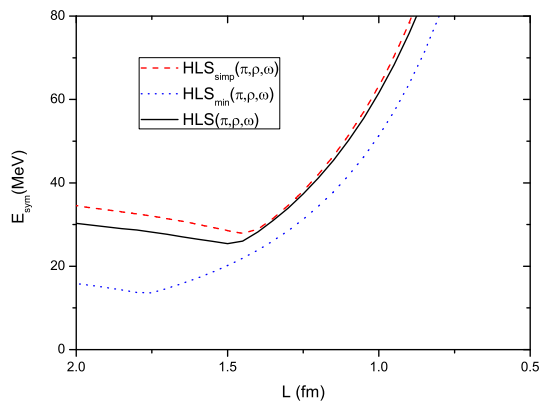}
\caption{The symmetry energy as a function of crystal size $L$ in HLS$(\pi,\rho,\omega)$ (solid line),  HLS$_{\rm min}(\pi,\rho,\omega)$ (dotted line) and  HLS$_{\rm simp}(\pi,\rho,\omega)$.}
\label{fig:EsymHLSComp}
\end{center}
\end{figure}

We compare the symmetry energy calculated with three versions of HLS model including $\pi, \rho$ and $\omega$ in Fig.~\ref{fig:EsymHLSComp}. From this figure we see that in both the skyrmion phase and half-skyrmion phase, the results from HLS$_{\rm min}(\pi,\rho,\omega)$ are significantly different from those of HLS$(\pi,\rho,\omega)$,  both in magnitude and in the location of $n_{1/2}$. As noted above, this is not surprising because HLS$_{\rm min}(\pi,\rho,\omega)$, with the $\rho$ mass taken to be infinite, does not follow the VM property as density increases. Thus it cannot be trusted at high density.  Truncating  the hWZ is a non-trivial matter, pointing at the crucial role of the $\omega$ in nuclear dynamics.

In contrast  HLS$_{\rm simp}(\pi,\rho,\omega)$ comes fairly close to HLS$(\pi,\rho,\omega)$. Above $n_{1/2}$, it is degenerate with the result of HLS$(\pi,\rho,\omega)$.
We noted above that the leading $O(p^2)$ HLS works remarkably well in the iso-vector sector.
This observation tells us that as in the static skyrmion matter~\cite{Ma:2016gdd}, the $z_4$ term of HLS Lagrangian~\eqref{eq:Lag_HLS} accounting for the $\rho$-$\pi$-$\pi$ interaction gives the dominant contribution from the normal parity $O(p^4)$ terms. It indicates that to study nucleon as well as  nuclear matter properties including both the $\rho$ meson and also $\omega$ meson effects, this simplified model will suffice, simplifying the calculation most dramatically.


\begin{figure}[h]
\begin{center}
\includegraphics[width=8cm]{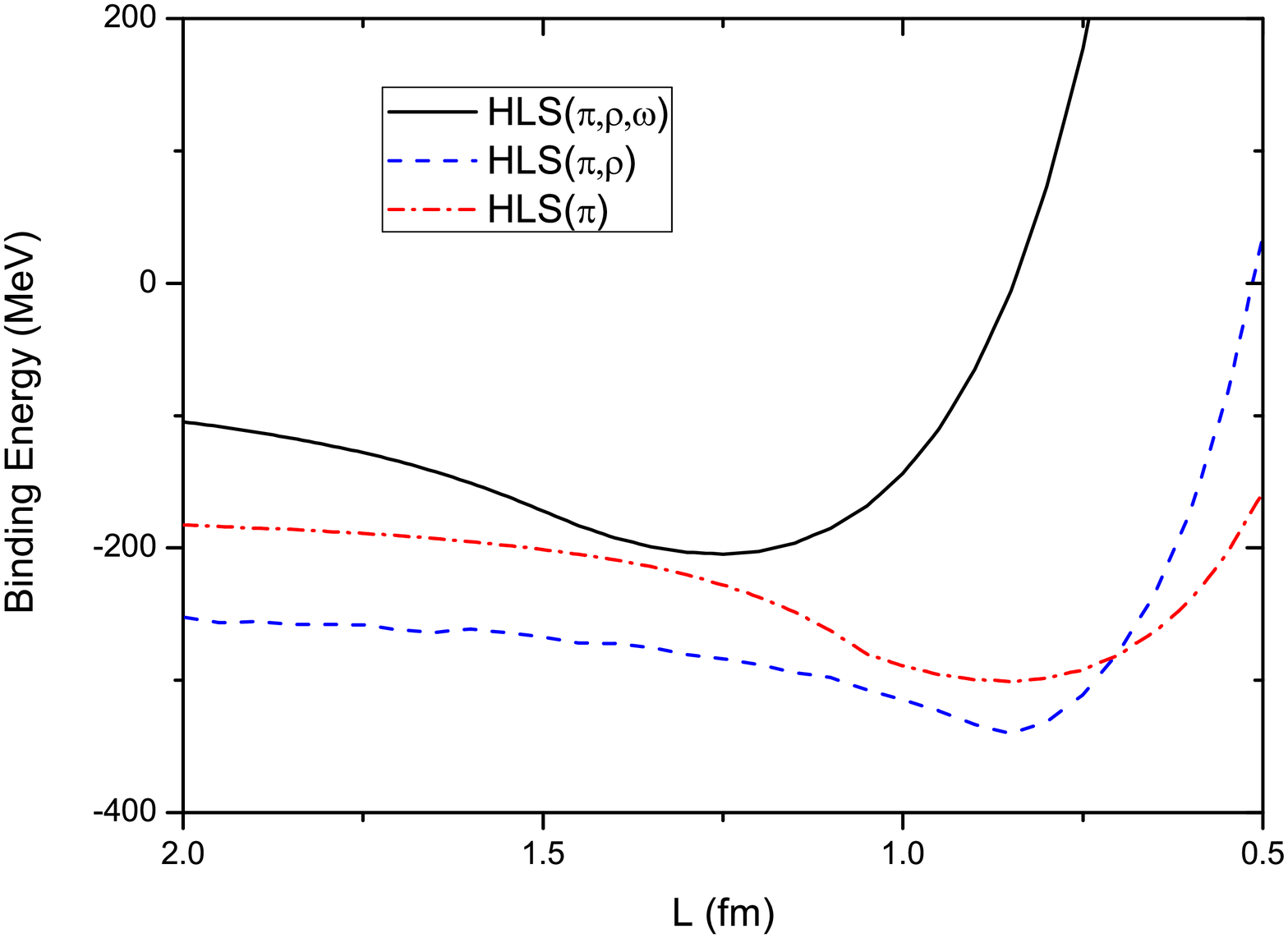}
\caption{Per neutron energy as a function of crystal size $L$ in HLS$(\pi,\rho,\omega)$ (solid line), HLS$(\pi,\rho)$ (dashed line) and HLS$(\pi)$ (dash-dotted line) with the corresponding calculated nucleon mass subtracted.}
\label{fig:PerMn}
\end{center}
\end{figure}

Although it is not our main concern in this paper, it is instructive to look at the ground-state energy per baryon obtained in the skyrmion crystal model. In QCD, calculating the ground-state energy from first-principle approaches has defied theorists since the beginning. On the basis of large $N_c$ considerations, one expects the scale of binding energy of many-nucleon systems to be order of { $\sim \Lambda_{QCD}$}, not a few MeV as observed in nature. At present we do not know how to explain satisfactorily the small binding energies of finite nuclei as well as the binding energy of nuclear matter $\sim 16$ MeV from QCD. It is interesting that  although it is not possible to compute the ground state of nuclear matter from first principles, it is however  found~\cite{sutcliffe-cluster} that putting the $\rho$ meson in the skyrmion description improves greatly in reducing binding energy of, as well as accounting for clustering phenomena observed in, light nuclei.

There is no way that the skyrmion crystal structure can give anything near reasonable for the ground state property\footnote{ In all works on skyrmion approaches to baryons and nuclear matter, it is overlooked that  the $O(N_c^0)$ terms, so-called Casimir terms, can be very important  but are ignored because they are daunting to calculate. For instance, the Casimir contribution to the baryon mass could be  of  $\sim 50\%$ of the leading  $O(N_c)$ term, much larger than the $O(1/N_c)$ term calculated by the collective-quantization, but comes with an opposite sign. In nuclear physics, it is the interplay between the scalar attraction due to a scalar meson $\sigma$ and the vector repulsion due to the vector meson $\omega$ that determines the ground state property of nuclear matter. The Casimir attraction could perhaps be simulated with an explicit scalar meson field in skyrmion approaches, but this has not been successfully worked out yet.}.  As one can see in Fig.~\ref{fig:PerMn}, all models give over-binding and too high saturation density. They are comparable to what is predicted by different approaches to skyrmion description. It is however interesting to see the effect of $\rho$ meson going in the right direction similarly to what's found in \cite{sutcliffe-cluster} and that the $\omega$ meson, not included in \cite{sutcliffe-cluster}, tends to lower the binding energy even further. This suggests that both vector mesons must be essential to understand the properties of nuclear matter in effective field theories.

In nuclear-structure physics, while the ground-state energy is extremely difficult to calculate, the energy of the excited states relative to the ground state is much easier to compute, in some cases rather accurately.  Since the symmetry energy is the energy difference between two states, $(E(n,\alpha=1)-E(n,\alpha=0))|_{n=\infty}$, one may hope to do better with it in the present model.  Indeed this is what one finds in Fig.~\ref{fig:EsymHLS}. While  one cannot trust the skymion lattice picture at low densities, say, far below $n_{1/2}$, it is amusing that the magnitude of the symmetry energy does come out to be $\sim (25 - 33)$~MeV  which is in the ball-park of experimental bounds.
\section{Conclusion and perspective}

\label{sec:dis}

In this paper, we approach the dense compact matter properties by collective-quantizing the skyrmion matter put on crystal lattice, described with the vector $\rho$ and $\omega$ mesons incorporated into the chiral Lagrangian via hidden local symmetry. We show that the cusp-like structure of the nuclear symmetry energy found in the Skyrme model -- with pion field supplemented with a scalar pseudo-dilaton $\chi$ -- remains more or less intact by the massive  vector excitation.  We are consequently led to the firm conclusion that the cusp structure is topological, mainly driven by the pion that carries topology. Being topological we believe it is highly robust.

It should of course be recognized that the cusp is only a leading-order effect. It is valid in the large $N_c$ limit from the  QCD side and in the large $\bar{N}$ limit  from nuclear correlations side. Higher-order effects together with explicit chiral symmetry breaking ignored in the numerical calculations should naturally smoothen the cusp as mentioned above. What remains as an imprint of that cusp in the EoS, which is visible as confirmed in a full $V_{lowk}$ RG flow,  is the changeover from ``softness" to ``hardness" in the EoS for compact stars accounting for the star mass $\gsim 2.0 M_\odot$ ~\cite{PKLMR} and also for the possible onset of a precocious pseudo-conformal sound velocity $v_s^2/c^2=1/3$ setting in for $n\gsim n_{1/2}$ in conjunction with the tidal deformability~\cite{vs-td}. These phenomena could perhaps be formulated as the emergence of the hidden symmetries in QCD, namely scale symmetry and flavor local symmetry. The changeover from skyrmions to half-skyrmions at $n_{1/2}$ could also be interpreted as a duality to the possible,  smooth, baryons-to-quarks transition. As they stand, they are conjectures to be vindicated by observations. Formulating the possible duality in terms of the Cheshire-Cat principle~\cite{cheshirecat} developed in 1980s would be a theoretical endeavor leading to a potential breakthrough.

If confirmed, this development would clear up the symmetry-energy landscape by weeding out the bulk of models that predict the symmetry energy decreasing to zero in the vicinity of $n\sim 3n_0$ in Fig.~\ref{mess}.

An important caveat in the treatment is that  the crystal treatment is undoubtedly invalid at low density,  most likely already at  $n_0$ and maybe even slightly below $n_{1/2}$ if not at $n\gsim n_{1/2}$.  Among other effects, the clustering of the type described in \cite{sutcliffe-cluster}  could play an important role just below $n_{1/2}$. Whether such an effect is adequately simulated in the crystal lattice with the inclusion of the vector mesons going beyond the collective quantization calls for a detailed further study.

Finally, the topology change phenomenon, even though buried in nuclear correlations, could affect the tidal deformability. Since the tidal deformability for star mass $M=1.4 M_\odot$ samples just below $n_{1/2}$, this is where the robust cusp structure figures. A more careful treatment of the topology change structure is in order here. This is an interesting open problem.

\subsection*{Acknowlegments}

Y.~L. Ma was supported in part by National Science Foundation of China (NSFC) under Grant No. 11475071, 11747308 and the Seeds Funding of Jilin University.

\appendix
\,

\section{Strategy of numerical simulation}

\label{App:}

In this appendix, we briefly describe how to simulate the symmetry energy or equivalently the momenta of inertia through Eq.~\eqref{eq:LEsymLambda} on crystal lattice directly from the Lagrangian.

To quantize the skyrmion crystal, we rotate the mesonic fields by a time-dependent angle, Eq.~\eqref{eq:rotation},  and define the rotation angle by using Eq.~\eqref{eq:defvelocity}. Then the Lagrangian of a certain chiral effective effective theory after the collective rotation can be written as
\begin{eqnarray}
{\cal L}_{\rm EFT} & = & {\cal L}_0 + {\cal L}(\bm{\Omega}),
\end{eqnarray}
where the first term ${\cal L}_0$ is the part independent of the velocity $\bm{\Omega}$ and ${\cal L}(\bm{\Omega})$ is the part including velocity. The moment of inertial is obtained from
\begin{eqnarray}
\frac{1}{4}\int_{\rm Cell} d^3 x {\cal L}(\bm{\Omega}) = \frac{1}{2} \lambda_I \bm{\Omega}^2,
\end{eqnarray}
where we have taken into account  that each cell of FCC crystal used in this work has 4 skyrmions.
Then, the moment of inertia of the system can be obtained as
\begin{eqnarray}
\lambda_I & = & \frac{1}{2\bm{\Omega}^2}\int_{\rm Cell} d^3 x {\cal L}(\bm{\Omega}) = \frac{1}{2}\int_{\rm Cell} d^3 x {\cal L}(\bm{\hat{\Omega}}).
\end{eqnarray}
where $\bm{\hat{\Omega}} = \bm{\Omega}/|\bm{\Omega}|$. As a result, the moment of inertia is a function of $\bm{\hat{\Omega}}$. The explicit form of  the moment of inertia $\lambda_I$ is quite involved. In this paper, we eschew deriving it and proceed to simulate $\lambda_I$  directly from the Lagrangian.

Briefly  the procedure is as follows.

In general, ${\cal L}(\bm{\hat{\Omega}})$ can be written as
\begin{eqnarray}
{\cal L}(\bm{\hat{\Omega}}) & = & A \left(\bm{\hat{\Omega}}\cdot \bm{\hat{\Omega}}\right) + \left(\bm{B}\cdot \bm{\hat{\Omega}}\right)\left(\bm{C}\cdot \bm{\hat{\Omega}}\right) \nonumber\\
& = &  A + \left(\bm{B}\cdot \bm{\hat{\Omega}}\right)\left(\bm{C}\cdot \bm{\hat{\Omega}}\right),
\end{eqnarray}
where we have used $\left(\bm{\hat{\Omega}}\cdot \bm{\hat{\Omega}}\right) = 1$. In doing the space integral in the crystal lattice, the angular part of the integral can be  approximated to
\begin{eqnarray}
\int \bm{\hat{\Omega}}_i \bm{\hat{\Omega}}_j \sin\theta d\theta d\phi & = & \frac{4\pi}{3}\delta_{ij}.
\end{eqnarray}
The moment of inertia takes the form
\begin{eqnarray}
\lambda_I 
& = & \frac{1}{3}\left[\frac{1}{2}\int_{\rm Cell} d^3 x \left(3A + \bm{B}\cdot \bm{C}\right)\right]\nonumber\\
& = & \frac{1}{3}\left[\left.\lambda_I\right|_{(1,0,0)} + \left.\lambda_I\right|_{(0,1,0)} + \left.\lambda_I\right|_{(0,0,1)}\right],
\end{eqnarray}
where
\begin{eqnarray}
\left.\lambda_I\right|_{(1,0,0)} & = & \frac{1}{2}\int_{\rm Cell} d^3 x \left(A + \bm{B}_1\bm{C}_1\right) \nonumber\\
\left.\lambda_I\right|_{(0,1,0)} & = & \frac{1}{2}\int_{\rm Cell} d^3 x \left(A + \bm{B}_2 \bm{C}_2\right)\nonumber\\
\left.\lambda_I\right|_{(0,0,1)} & = & \frac{1}{2}\int_{\rm Cell} d^3 x \left(A + \bm{B}_3 \bm{C}_3\right).
\end{eqnarray}
Since the crystal lattice is symmetric with respect to the $x, y$ and $z$ axis, one obtains
\begin{eqnarray}
\lambda_I & = & \left.\lambda_I\right|_{(1,0,0)} = \left.\lambda_I\right|_{(0,1,0)} = \left.\lambda_I\right|_{(0,0,1)}.
\end{eqnarray}
Therefore, in the simulation, one can specify a direction of $\bm{\Omega}$, for example $\bm{\Omega}_1$, to calculate the moment of inertia.

For a typical example, consider the second term in the $O(p^2)$ HLS
\begin{eqnarray}
{\cal L}_{(2)}^2 & = & a f_\pi^2 \,\mbox{Tr}\, \left(\hat{a}_{\parallel\mu}
\hat{a}_{\parallel}^{\mu} \right),
\end{eqnarray}
which includes all the degrees of freedom, $\pi$,  $\rho$ and $\omega$.  Then, the contribution to the moment of inertia can be written as
\begin{eqnarray}
{\cal L}_{(2)}^2 & = & a f_\pi^2 \,\mbox{Tr}\, \left(\hat{a}_{\parallel 0}^{(3)}
\hat{a}_{\parallel 0}^{(3)} \right) - a f_\pi^2 \,\mbox{Tr}\, \left(\bm{\omega}_i\bm{\omega}_i\right) ,
\label{Eq:Op2T2}
\end{eqnarray}
where $\hat{a}_{\parallel 0}^{(3)}$ is the $SU(2)$ part of $\hat{a}_{\parallel 0}$. After choosing the specific direction of the angular velocity as, for example $x$-direction, the basic quantities in \eqref{Eq:Op2T2} an be written as
\begin{eqnarray}
\partial_0 \xi & = & \frac{i}{2}C \left[ \bm{\tau}_1, \xi_c\right] C^\dagger,\nonumber\\
V_0 & = & \frac{1}{1+ \bar{ \phi} _0^{\rho_ 0}}C(t)\left[\left(\bm{\bar{\phi}}^{\rho_0}\cdot\bm{\bar{\phi}}^{\rho_0}\right)\bm{\tau}_1 - \bm{\bar{\phi}}^{\rho_0}_1 \left(\bm{\tau}\cdot \bm{\bar{\phi}}^{\rho_0}\right) \right]C^\dagger(t),\nonumber\\
\bm{\omega}_i & = & \epsilon_{i1k}\bm{\omega}_k.
\end{eqnarray}
Substituting these expressions into \eqref{Eq:Op2T2}, taking the trace of Pauli matrices and  minimizing the momenta of inertia of the full theory by adjusting the Fourier coefficients of $\bar{\bm{\phi}}^{\rho_0}$ and $\bm{\omega}$ with $\xi_c$ determined from the simulation of the energy per baryon in the static crystal, one obtains the moment of inertia.

\end{document}